# Influence of external magnetic field on laser breakdown plasma in aqueous Au nanoparticles colloidal solutions

A.A. Serkov[1,2*], I.I. Rakov[1], A.V. Simakin[1], P.G. Kuzmin[1], G.N. Mikhailova[3], L.Kh. Antonova[3], A.V. Troitskii[3], G.P. Kuzmin[3], and G.A. Shafeev[1,4]

[1]Wave Research Center of A.M. Prokhorov General Physics Institute of the Russian Academy of Sciences, 38, Vavilov street, 119991 Moscow Russian Federation

[2]The Federal State Educational Institution of Higher Professional Education, "Moscow Institute of Physics and Technology (State University)", Moscow, Russian Federation

[3]A.M. Prokhorov General Physics Institute of the Russian Academy of Sciences, 38, Vavilov street, 119991 Moscow Russian Federation

[4]National Research Nuclear University MEPhI (Moscow Engineering Physics Institute), 31, Kashirskoye highway, 115409, Moscow, Russian Federation

*antonserkov@gmail.com

## Abstract

Influence of permanent magnetic field up to 7.5 T on plasma emission and laser-assisted Au nanoparticles fragmentation in water is experimentally studied. It is found that presence of magnetic field causes the breakdown plasma emission to start earlier regarding to laser pulse. Field presence also accelerates the fragmentation of nanoparticles down to a few nanometers. Dependence of Au NPs fragmentation rate in water on magnetic field intensity is investigated. The results are discussed on the basis of laser-induced plasma interaction with magnetic field.

## Introduction

The peculiar feature of infrared laser radiation interaction with NP colloidal solutions consists in plasma excitation around individual NPs inside the beam waist during the pulse. Indeed, plasmon resonances of most metallic NPs are situated either in the visible or UV range of spectrum [1]. Therefore, the efficient cross section of laser radiation absorption by NPs is close to their geometric cross section [2]. If laser intensity is high enough for NPs to reach temperatures of about $10^4$-$10^5$ K, some part of their atoms may be ionized [3]. When the fraction of these atoms reaches the critical value, plasma formation occurs [4]. Such type of plasma is often referred to as "nanoplasma" [5] due to its confinement in the small region in the NP vicinity. One may assign its own plasmon resonance to this local plasma, which is determined by the concentration of electrons in it. At high laser pulse repetition rates these small sources of plasma may unite between each other resulting in the liquid breakdown plasma formation [6].

Exact mechanisms determining the process of the interaction of laser radiation with the NPs colloidal solutions are still under consideration. However, it is usually accepted that laser exposure of sufficient intensity leads to the decrease of the NPs mean size, i.e. their "fragmentation". Under different experimental conditions this phenomenon may be caused by different processes, such as hydrodynamic instabilities [7], [8] or Coulomb explosion [3], [9], [10]. One should note that use of the nanosecond laser pulses of the infrared wavelengths is of particular interest. Namely, due to the aforementioned phenomenon of plasma formation the radiation absorption is determined by both properties of the NP material and of the plasma. As it usually takes several nanoseconds for the plasma plume to emerge, it starts existing during the laser pulse, thus absorbing some part of it [11]. One of the possible fragmentation mechanisms in this case is condensation from the plasma plume consisting of both initial NP material and working liquid ions. The process begins with the nuclei formation and is followed by their aggregation at the plasma-working liquid interface. Properties of the NP produced this way are determined by both liquid and plasma plume properties [12]. Thus, one may propose a method to control these properties via influence on laser plasma. In earlier works it was shown that application of electric field during laser ablation from Ge target may affect the resulting NP morphology and chemical composition via interaction with the plasma plume [13]. The effect of external magnetic field on morphology and magnetization of iron alloys NPs produced via laser ablation was studied in [14]. In this work, however, the field was considered to affect the interaction forces between the NP due to their ferromagnetic properties.

The aim of the current work is the investigation of the effect of external magnetic field on laser-induced liquid breakdown and on the process of laser-assisted diamagnetic (Au) NPs fragmentation. It is shown that NP fragmentation rate, as well as the position of size distribution maximum depends on the external magnetic field intensity. The size distributions obtained by disk centrifuge are corroborated by Transmission Electron Microscopy (TEM) data, as well as absorption spectra of colloidal solutions. The discussion of the results is based on the interaction of the magnetic field with laser-induced plasma.

**Experimental technique**

Colloidal solutions of individual NPs were prepared using the technique of laser ablation in liquid [15]. For this purpose, an Ytterbium fiber laser with pulse duration of 70 ns, repetition rate of 20 kHz and pulse energy of 1 mJ at 1060-1070 nm was employed as irradiation source for NPs generation. Laser radiation was focused on metallic plate made of corresponding metal immersed into deionized water by an F-Theta objective. Laser beam was scanned across the sample surface at the speed of 1000 mm/s by means of galvo mirror system. Resulting concentration of NPs in colloidal solution was about $10^{14}$-$10^{15}$ particles/cm$^3$. This estimation was based on the change of mass of the metal target and NPs size distribution function [16]. Typical NPs generation rate is of about 0.5 mg/min. Some amount of Polyvinylpyrrolidone (PVP, molecular mass of $10^4$, about 0.5 mg/ml) was used in all experiments for NPs colloidal solution stabilization. PVP was added to working liquid before NPs generation. Size distribution of obtained NPs was analyzed with either Measuring Disc Centrifuge (CPS Instruments) (0.1 ml samples) or Transmission Electron Microscope (TEM) images of NPs.

Further irradiation of NPs colloidal solutions in absence of the target was carried out using the radiation of the Nd:YAG laser system Sol (Bright Solutions) at wavelength of 1064 nm and pulse duration of about 10 ns (FWHM). Laser radiation was focused into colloid through the transparent bottom of the cylindrical cell using a lens with focal distance of 25 mm. Laser exposure of 2 ml portions of colloids was carried out at 2 mJ energy per pulse and repetition rate of laser pulses of 10 kHz. The magnetic field was provided by means of the cryogen-free magnetic system with a refrigerator. The superconducting solenoid from Nb-Ti alloy wire provided the magnetic field up to 8 T [17]. The cell was fixed in the middle part of the magnet where the field is homogeneous. Experimental setup on laser irradiation of NPs colloidal solution in external magnetic field is presented in Fig.1. Plasma emission was collected using an optical fiber placed right above the cell via the collimator lens and then was directed onto a pin-diode. The cut-off glass filter was placed between the cell and optical fiber in order to reduce the transmitted laser radiation.

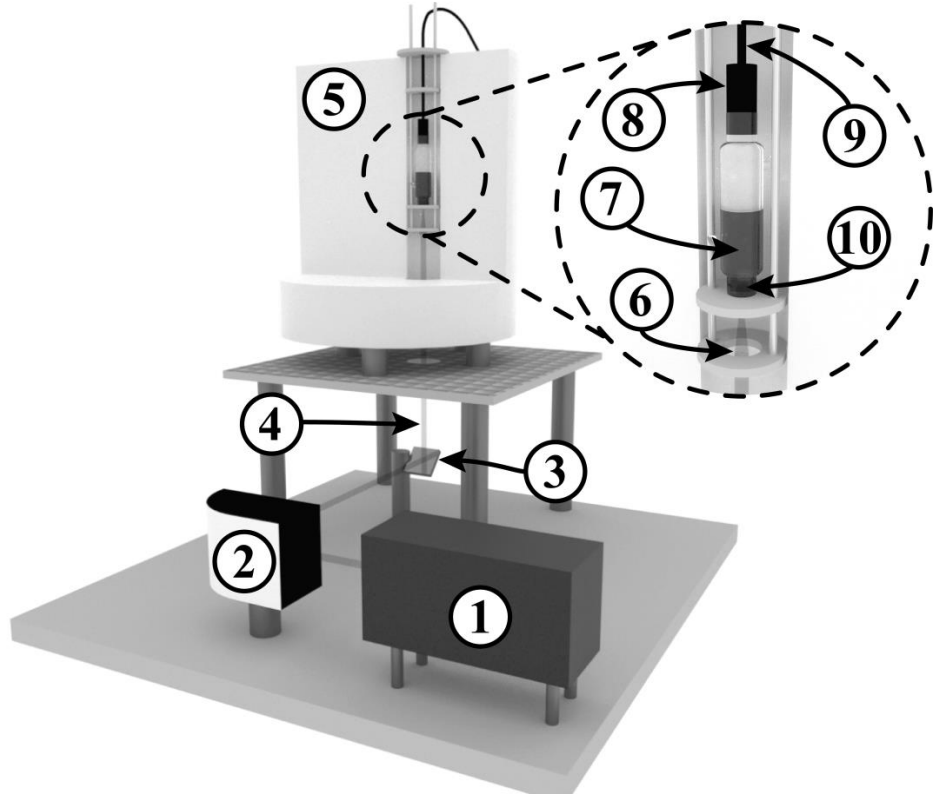

Fig.1. Experimental setup on laser fragmentation in external magnetic field. Laser source (1), galvo mirror system (2), dielectric mirror (3), laser beam (4), magnet (5), focusing lens (6), NP colloid (7), fiber collimator lens (8), optic fiber (9), laser beam waist (10)

The morphology of metallic NPs was analyzed by means of Transmission Electron Microscopy (TEM). Extinction spectra of colloidal solutions were acquired using an Ocean Optics UV-Vis fiber spectrometer in the range of 250 - 800 nm.

**Results and Discussion**

In order to investigate the effect of external magnetic field on laser plasma we studied the dependencies of its spectral and temporal characteristics on the field intensity. Typical spectrum of laser plasma in liquid is continuous in the visible area due to recombination and Bremsstrahlung emission [18] (Fig. 2).

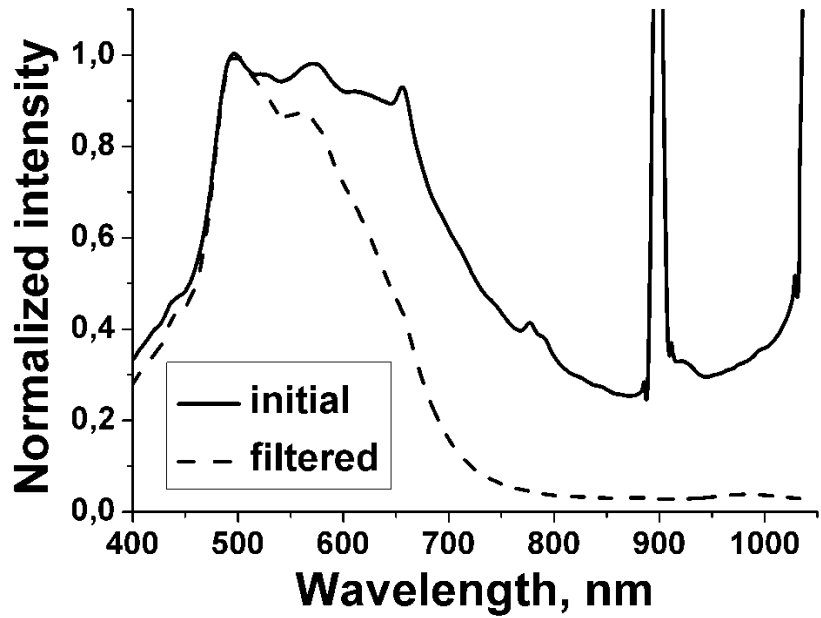

Fig. 2. Typical spectrum of laser plasma in gold NPs aqueous colloidal solution (black line, without cut-off filter), spectrum after the filter (dashed line). Nd:YAG laser, wavelength of 1064 nm, pulse duration of 10 ns, pulse energy of 3 mJ

According to the spectral measurements, introduction of external field causes almost no change in laser plasma spectrum. It should also be noted that in order to measure the temporal characteristics of the plasma we used the filtered signal (dashed line in Fig. 2). Typical oscillograms of both plasma emission and laser pulse are presented in Fig. 3.

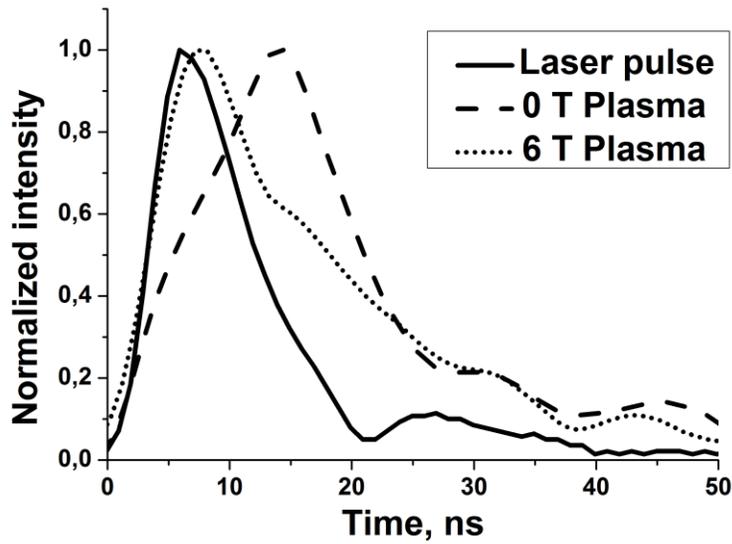

Fig. 3. Typical oscillograms of laser pulse and laser plasma emission. Nd:YAG laser radiation in gold NPs aqueous colloidal solution, wavelength of 1064 nm, pulse duration of 10 ns, pulse energy of 3 mJ

As one can see, application of external magnetic field significantly alters the temporal characteristics of the laser plasma signal. Namely, both the shape and delay between the laser pulse and plasma emission are changed.

The temporal shape of the signal is considerably asymmetrical. Thus, in order to analyze the impact of the field on shape of the laser plasma signal, we studied the dependency of the integral plasma signal (normalized to integral laser pulse signal) on the field intensity. The impact of magnetic field on delay between the maxima of laser pulse and plasma signal was also analyzed. The resulting data are presented in Fig. 4a and 4b.

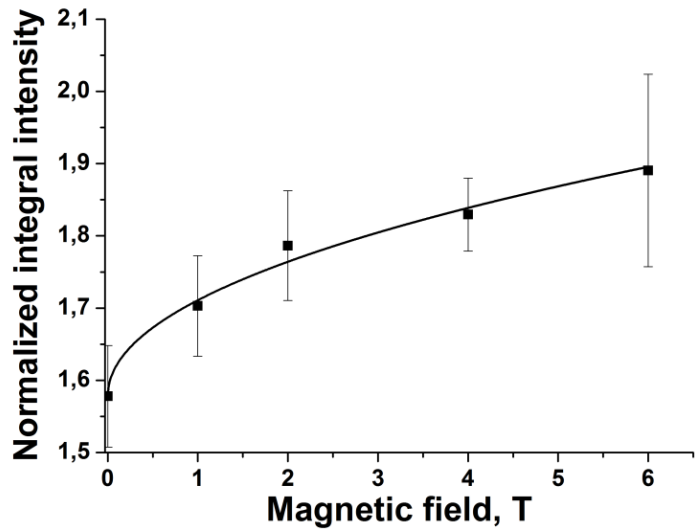

a

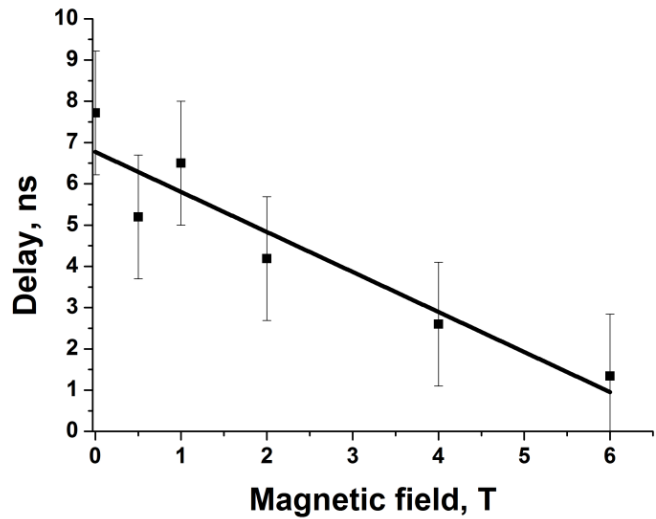

b

Fig. 4. Dependencies of the normalized plasma intensity (a) and delay between the maxima of the laser ulse and plasma emission signal (b) on magnetic field intensity

According to the plasma signal measurements, the integral intensity shows a 20% increase after an increase of the field intensity from 0 to 6 Tesla. The delay between the maxima of laser pulse and plasma signals also shows a clear dependence on field intensity. As one can see in Fig. 4b, without external magnetic field delay reaches the value of 8 ns, while at 6 T it plummets down to about 1.5 ns. Using these experimental results, one can conclude that presence of external magnetic field alters the breakdown threshold of laser-induced plasma and leads to an increase in its emission intensity.

In order to investigate the influence of external magnetic field on fragmentation process several series of experiments under different conditions were carried out. First of them included exposure of Au NPs colloidal solutions to nanosecond laser radiation with and without magnetic field (constant intensity of 7.5 T). Exposure time was varied in the range of 5 and 30 minutes. Morphology of the initial Au NPs and of the NPs after 30 minutes exposure in magnetic field is shown in Fig. 5.

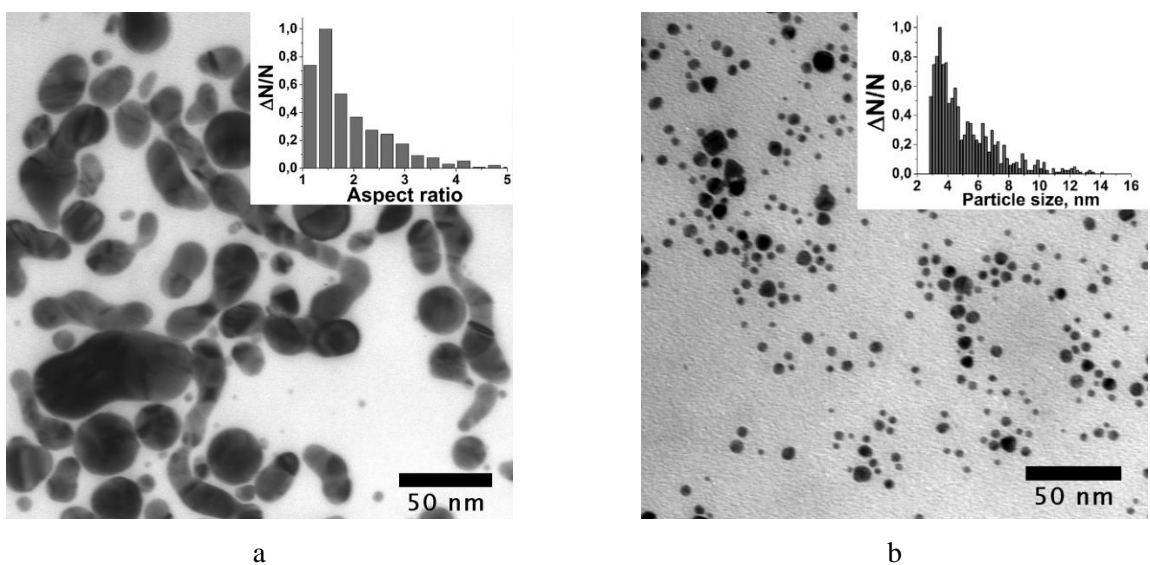

a  b

Fig. 5. TEM images of initial Au NPs (a) and Au NPs after 30 min exposure to nanosecond laser radiation in 7.5 T magnetic field (b). The insets show aspect ratio (AR) (a) and size (b) distributions

According to Fig. 5, the initial colloidal solution consists of elongated NPs with transverse size of several tens of nanometers and aspect ratio (AR) ranging from 1 to 4. Laser exposure in magnetic field leads to significant decrease in NPs size and change of NPs shape to spherical. Thus, external magnetic field causes no change in NPs morphology compared with fragmentation without field [19], [20]. However, the size of resulting NPs obtained at the same exposure time in magnetic field is much lower than without it.

Evolution of Au NPs size distribution functions measured by disk centrifuge is presented in Fig.6.

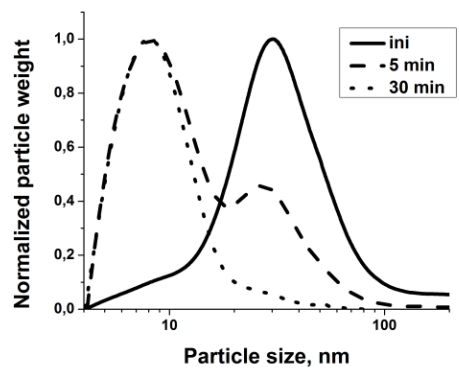

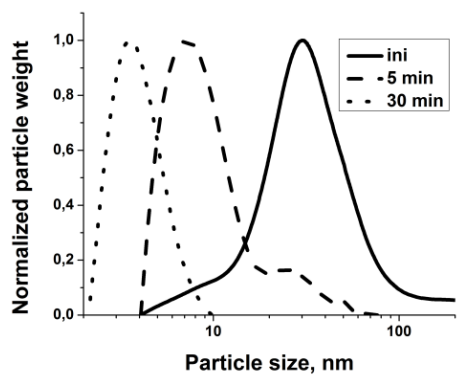

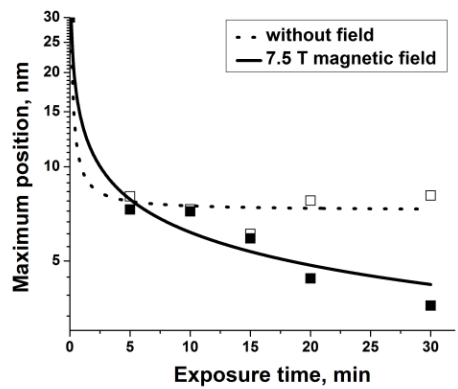

Fig.6. Evolution of size distribution functions of Au NPs without external field (a), in external magnetic field of 7.5 T (b). Effect of the magnetic field on the size distribution maximum position (c). Hollow square markers denote the maximum position without magnetic field, whereas the black ones account for the maximum in 7.5 T magnetic field position

As one can see in Fig.6, external magnetic field affects fragmentation. Magnetic field also causes gold NPs size distribution function maximum to shift towards smaller sizes. Without magnetic field the value of size distribution maximum amounts to 8 nm, while introduction of field causes it to reach almost 3 nm at the same time of laser exposure.

According to the size distributions in Fig.6, under given conditions the main fraction of the initial NPs become fragmented in the first 5 minutes of laser exposure. To investigate the process more thoroughly a series of experiments with the exposure times of 1, 2, 4, and 6 minutes was carried out. Magnetic field intensity was varied from 2 to 7.5 T. Typical dependencies of size distribution on field intensity after 1 min exposure are presented in Fig. 7.

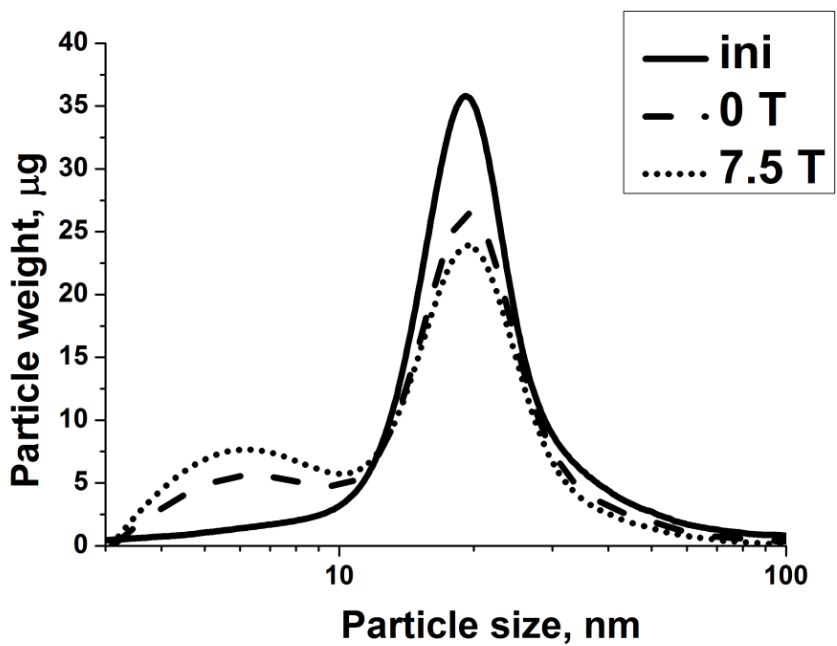

Fig. 7. Size distribution functions dependency on magnetic field intensity for Au NPs. Exposure time of 1 min

The size distributions given in Fig.7 show that application of external magnetic field does not affect the position of additional maximum at size of 6 nm. The time dependency of maximum amplitude at this size for different values of magnetic field intensity are presented in Fig. 8.

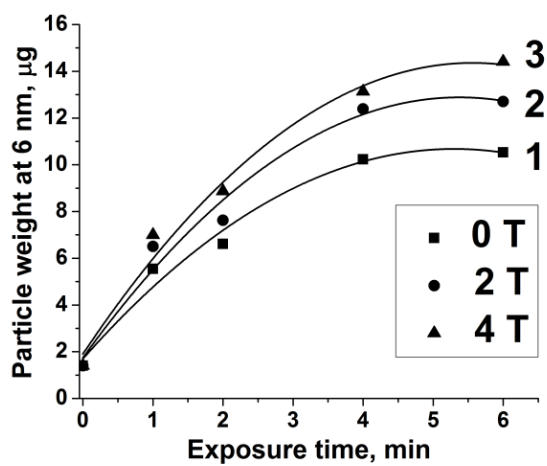

Fig. 8. Time dependencies of small-size maximum amplitude of Au NPs. Curves 1, 2, and 3 account for magnetic field intensities of 0, 2, and 4 Tesla, respectively.

According to Fig. 8, the amplitudes of small-size maximum show a steady growth with the increase of exposure time, which is typical for fragmentation process [21]. Magnetic field causes additional increase of 6 nm Au NPs number. The dependence of the amplitude on magnetic field intensity is also clearly visible. It can be seen that magnetic field of 4 T field intensity causes the 1.4 times increase of 6 nm Au NP weight (from 10.4 to 14.4 µg) after 6 minutes exposure.

The disk centrifuge measurements are corroborated by extinction spectra of colloidal solutions (Fig. 9)

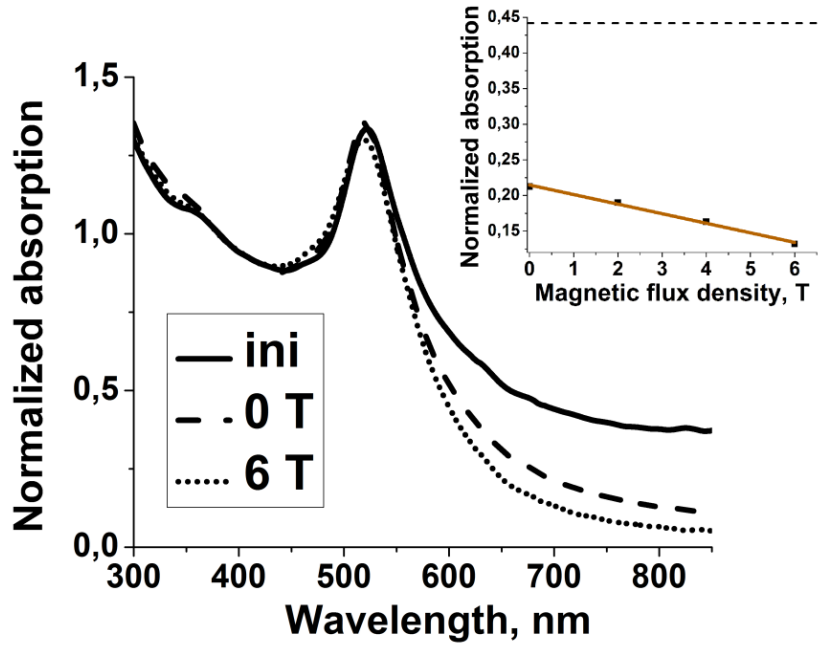

Fig.9. Normalized Au NPs colloidal solutions absorption spectra after 2 minutes exposure. The inset shows the dependence of normalized absorption at wavelength of 700 nm on magnetic field intensity. The dashed line in the inset accounts for absorption at 700 nm of the initial colloidal solution

Initial spectrum of the colloidal solution (ini) contains wide red wing typical of elongated Au NPs shown in Fig. 5 a ( [22]). As one can see in Fig.9, optical density of the colloidal solution in the red region decreases with the increase of magnetic field intensity. The decrease of absorbance in this spectral region, therefore, corresponds to elongated NPs fragmentation.

The influence of magnetic field on laser-induced NPs fragmentation may manifest itself in two different ways. First of them is the direct interaction of the field with the NPs dispersed in liquid [23]. However, that scenario would be more probable for the ferromagnetic materials, such as iron and cobalt. Although gold NPs may exhibit some magnetic properties [24], most of them occur on single-particle scale and bear the paramagnetic character. Another possible mechanism is the indirect one and is mediated through laser-induced plasma. Indeed, laser intensity used in current experiments ($10^{10}$ W/cm$^2$) is comparable to the value of laser-induced breakdown threshold intensity in pure water [18]. Given that presence of nanoparticles in water can significantly decrease this threshold intensity [25], we can conclude that laser-induced breakdown indeed takes place in our experiments. Moreover, we can evaluate the plasma parameters according to nanoparticle atoms ionization process.

If NP temperature reaches the value of approximately $10^4$-$10^5$ K, it may be partially ionized [3]. Let us evaluate this temperature based on the model presented in [21]. One should note that according to this model NPs absorbance is governed by that of bulk gold.

NP temperature can be estimated as:

$$T = T_0 \exp\left(\frac{R^2}{R_0^2}\right) \quad (1)$$

where $T_0$ is room temperature (300 K), $R$ is the NP radius, and $R_0^2$ can be presented as:

$$R_0^2 = \frac{\kappa_0 T_0 \lambda}{2\pi I} \times \frac{1}{Im\left(\frac{\varepsilon - \varepsilon_0}{\varepsilon + 2\varepsilon_0}\right)} \quad (2)$$

where $\kappa_0$ is thermal conductivity of water (0.6 W×m$^{-1}$×K$^{-1}$), $\lambda$ is wavelength of incident laser radiation (1064 nm), I is intensity of incident laser radiation (equal to $10^{10}$ W/cm$^2$ at peak power of $2\times10^5$ W), $\varepsilon$ is dielectric constant of gold at 1064 nm (-43.835+4.1736$i$), $\varepsilon_0$ is dielectric constant of surrounding medium at 1064 nm (1.7585). Using these parameters, we can calculate $R_0^2$ to be equal to 140 nm$^2$. Therefore, the 30 nm NP (Fig.5 a) temperature will be approximately equal to $2\times10^5$ K. The number of electrons ejected from that NP via thermionic emission can be evaluated using the Richardson-Dushmann equation [26]:

$$j = A_0(1-R) \times T^2 \times \exp\left(\frac{-e\phi}{kT}\right) \quad (3)$$

where $\phi = 5.1\ eV$ is the work function of gold, $T = 2 \times 10^5\ K$ is the NP temperature, and $e$ is electron charge. The electron emission efficiency, thus, is proportional to $\exp\left(\frac{-\phi}{kT}\right)$. Using the given parameters we can evaluate that the fraction of emitted electrons is sufficient for generation of nanoplasma in the vicinity of NPs [5] which eventually leads to laser-induced breakdown [27], [6].

The effect of strong external magnetic field on the threshold of laser-induced breakdown is a well-known phenomenon for the gas media [28], [29]. In case of liquid media it is usually discussed on the basis of its impact on the laser-induced breakdown spectroscopy (LIBS). Namely, authors of [30] have shown that external field can lead to intensity enhancement of the plasma signal, thus increasing the sensitivity of the method. The effect of the external magnetic field was also studied in case of the laser plasma produced on the surface of a target immersed in liquid. As it was reported in [31], [32], in this case external magnetic field can significantly increase the lifetime of laser induced plasma.

The impact of the field on laser-induced plasma in current work can be estimated via the crucial parameters that determine plasma behavior in external magnetic field. One of them is electron Larmor radius of those electrons that have the component of velocity perpendicular to magnetic field. It can be evaluated based on the fact that average electron energy equals to [5]:

$$\varepsilon_{av} = \frac{E_{ion}}{2} \qquad (4)$$

where $E_{ion}$ is the electron binding energy (equal to work function of gold, 5.1 eV). In such a way we can obtain electron Larmor radius equal to $r_L \approx 9 \times 10^{-7}\ m$ for magnetic field of 6 T. Due to the fact that laser-induced breakdown occurs only when the threshold density of free electrons is reached [18], we can assume that external field affects the diffusion speed of the electrons from the laser beam waist region [33]. Namely, the width of this area is of about *$10^{-5}$ m* whereas the Larmor radius equals to ≈*$10^{-6}$ m*.

Another important parameter is the so-called "bouncing radius" of plasma. Namely, it defines the size of the sphere, in which plasma in external magnetic field is confined [34]. We can evaluate it as follows:

$$R_b = \sqrt[3]{\frac{3E_p\mu_0}{4\pi B^2}} \qquad (5)$$

where $E_p$ is the laser plasma energy, $\mu_0$ is magnetic permeability of vacuum, $B$ is the applied magnetic field. One should note that we estimate the value of the plasma energy to be approximately equal to the laser pulse energy [35]. In reality this value may be lower; however we consider that approximation to be legitimate in our case. Thus, we calculate the $R_b$ to be equal to $\approx 0.2 \times 10^{-3}\ m$. One can now compare this value to the size of the cavitation bubble, in which laser-induced plasma in liquid is confined. According to the point explosion theory [35], [36], it can be defined as:

$$R = \left(\frac{5}{2}\right)^{\frac{2}{5}} \left(\frac{E_p}{2\pi\rho}\right)^{\frac{1}{5}} t^{\frac{2}{5}} \qquad (6)$$

where $R$ is the bubble radius, $E_p$ is the laser plasma energy, $\rho$ is density of water, $t$ corresponds to time. According to the experimental data, the lifetime of plasma is of about 10-30 ns. Using these parameters, we evaluate $R$ to be approximately equal to 50-100 µm.

As one can see, the values of the cavitation bubble and bouncing radii are of the same order. Thus, on one hand, the laser-induced plasma plume is confined by the liquid itself. On the other hand, external magnetic field introduces another mechanism of confinement. Although in current work we could not precisely measure the values of the radii, the aforementioned evaluations lead us to conclusion that the intensity of the magnetic field used was enough to effectively confine the plasma plume. Therefore, according to the model given in [37], the increased emission of the plasma plume can be accounted for enhanced recombination and Bremsstrahlung.

The increased plasma emission, however, does not account for the enhanced NP fragmentation observed in our experiments. According to the energy balance given in [35], less than 1 percent of the incoming laser radiation is converted into plasma radiation. The effect of this emission on NP temperature can be evaluated using (2). Since plasma emission can be approximated as a continuum in the visible range, one needs to integrate $\frac{\lambda}{Im(\frac{\varepsilon-\varepsilon_0}{\varepsilon+2\varepsilon_0})}$, which corresponds to NP absorption at given wavelength. The resulting value of 30 nm NP temperature is of approximately 330 K at initial temperature of 300 K. This fact leads us to a conclusion that the laser plasma emission does not significantly affect the NP fragmentation process.

Another factor that may affect the fragmentation process is the increased laser plasma temperature. As mentioned above, laser-induced breakdown threshold in external magnetic field is decreased due to the altered diffusion velocity of electrons. This fact, along with the consequent plasma confinement, results in generation of denser plasma with increased lifetime. That plasma, in turn, can be characterized by increased electron collision rate and increased electron temperature [29].

Due to the fact that on the earlier stages laser plasma is formed around NPs, we assume the heat transfer from plasma to NP to be efficient. Exact mechanisms of plasma influence on NP fragmentation, however, require further investigation. We consider the most probable scenario to be the NP temperature increase due to heat transfer from plasma followed by the NP fragmentation via Coulomb explosion [10] or hydrodynamic instabilities [16].

On the other hand, the NP formation may be considered as a result of material condensation from plasma plume. In such a case NP size is defined by the critical cluster size equal to [12]:

$$n_c = \left[\frac{8\pi a^2 \sigma}{3kTln(S)}\right]^3 \qquad (7)$$

where $a$ is the effective radius of gold atom, $\sigma$ is the effective surface tension, $k$ is the Boltzmann constant, $T$ is the plasma temperature, $S$ is the super saturation parameter. As one can see, the increase of the plume temperature will strongly affect the critical cluster size.

**Conclusion**

Influence of external magnetic field on laser assisted breakdown plasma in aqueous Au nanoparticles colloidal solutions was studied. The field of 6 T was shown to effectively confine the plasma plume leading to increased lifetime and enhanced emission intensity. Possible mechanisms of energy transfer from plasma plume to gold NPs were discussed. It was shown that laser-induced fragmentation of gold NPs dispersed in water in external magnetic is accelerated, resulting in smaller sizes of final NPs.

**Acknowledgments:**

The authors gratefully acknowledge the support of the RF President's Grants Council for Support of Leading Scientific Schools (Grant # NSh-4484.2014.2), RFBR Grants #15-02-04510 A, #16-02-01054, #15-32-20926 and the Presidential scholarship for young scientists and postgraduate students SP-753.2015.2.